\documentclass[12pt]{iopart}

\usepackage{iopams}
\usepackage{graphicx}
\begin{document}
\title[The anisotropic flow coefficients $v_2$ and $v_4$ in Au+Au collisions at RHIC]
{The anisotropic flow coefficients $v_2$ and $v_4$ in Au+Au collisions at RHIC}

\author{Yuting Bai for the STAR\footnote{For the full list of STAR authors and acknowledgements, 
see appendix `Collaborations' in this volume.} Collaboration}
\address{NIKHEF, Kruislaan 409, 1098SJ Amsterdam, The Netherlands}
\ead{Yuting.Bai@nikhef.nl}
\begin{abstract}
We present measurements by STAR of the anisotropic flow coefficients $v_2$ and $v_4$ as a
function of particle-type, centrality, transverse momentum and pseudorapidity in
Au+Au collisions at RHIC.
\end{abstract}

Anisotropic flow is an azimuthal correlation of the particle momenta
with respect to the reaction plane. This flow is recognized as one of
the main observables that provide information on the early stage of a
heavy-ion collision~\cite{Whitepaper}.

In this study, we used $13 \times 10^6$ minimum-bias Au+Au events taken at 
a center-of-mass energy of 200~GeV and $6 \times 10^6$ events taken at 62.4~GeV.
The particles
were detected by the STAR main TPC~\cite{TPC} and by the forward
TPCs~\cite{FTPC} and cover a pseudorapidity of $|\eta|<1.3$ and
$2.5<|\eta|<4.0$, respectively. From these data, $v_2$ is obtained
by the 4-particle cumulant method~\cite{Cumu} and is denoted by
$v_2\{4\}$. This method is less sensitive to non-flow effects compared
to measurements based on two particle correlations like $v_2\{2\}$ or
$v_2\{EP_2\}$. The $v_4$ coefficient is obtained with respect to the second
harmonic event plane and is denoted by $v_4\{EP_2\}$. The flow coefficients
are studied for different particle species as function of transverse momentum ($p_t$),
pseudorapidity ($\eta$) and centrality. Only statistical errors are shown unless specified.

\begin{figure}
\centering
\begin{minipage}[t]{66mm}
\includegraphics[width=1.0\textwidth]{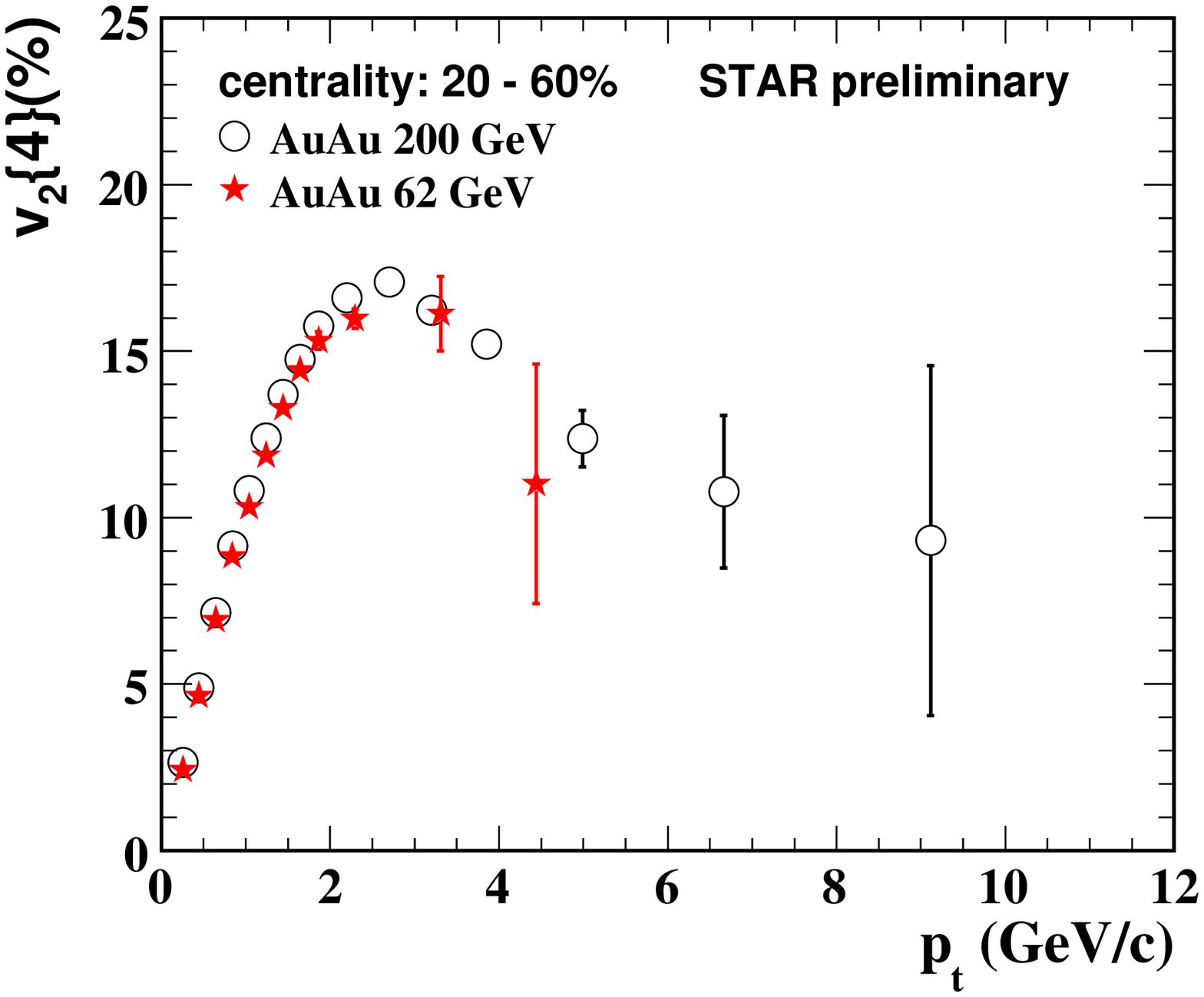}
\end{minipage}
\begin{minipage}[t]{66mm}
\includegraphics[width=1.0\textwidth]{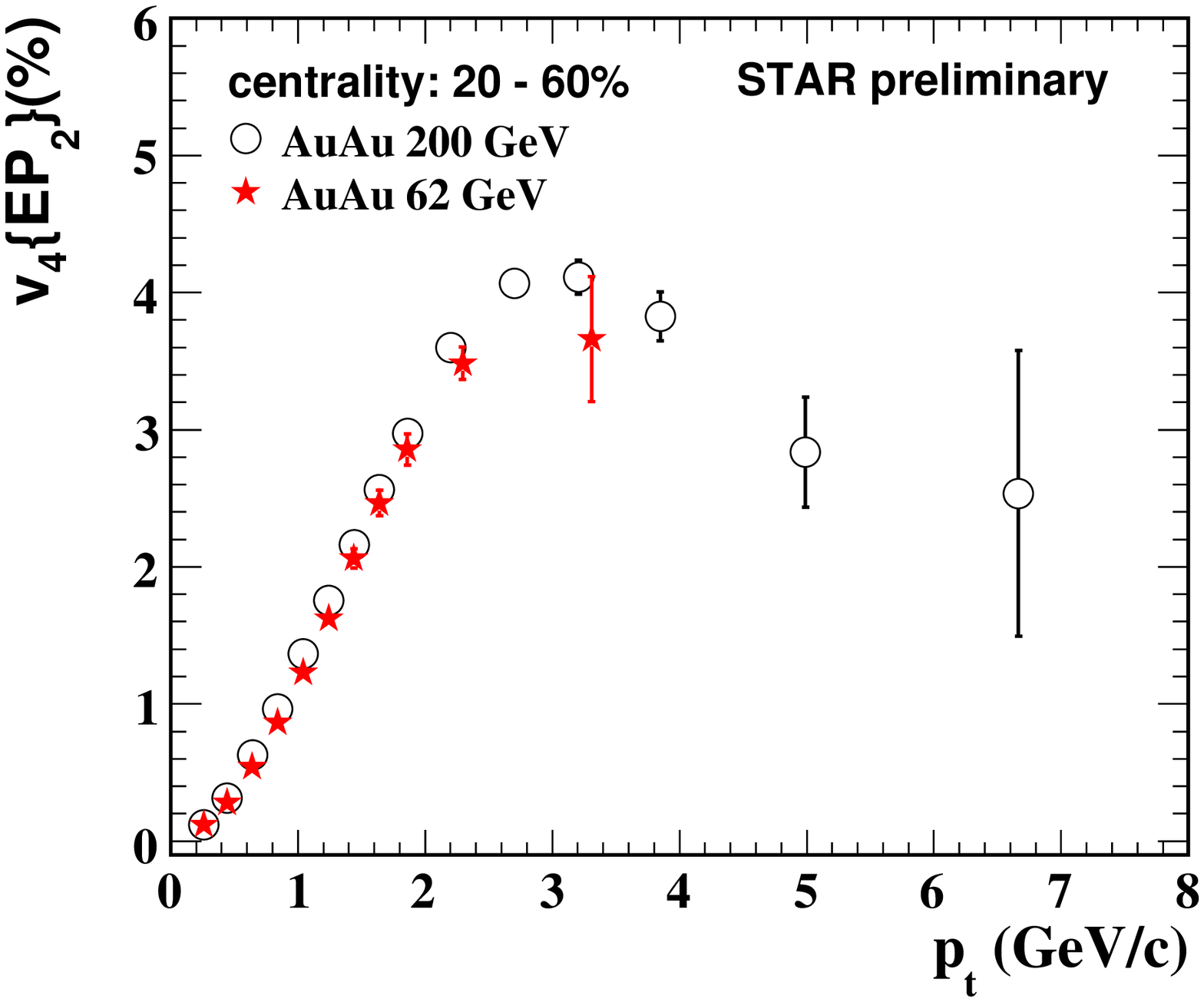}
\end{minipage}
\caption{Charged particle $v_2\{4\}$ (left-hand plot) and $v_4\{EP_2\}$ (right-hand plot) 
versus $p_t$ for 20--60\% centrality at $|\eta| < 1.3$ in Au + Au collisions at 200
and 62.4~GeV.}
\label{fig:v2_pt}
\end{figure}

\paragraph{Transverse momentum dependence.}
The left panel of Fig.~\ref{fig:v2_pt} shows the charged particle
$v_2$ as a function of $p_t$ for mid-central (20--60\%) Au+Au
collisions at 200 and 62.4~GeV. It is seen that the
measured $v_2$ increases with $p_t$, reaches its maximum around
3~GeV/$c$ and then decreases again. At 200~GeV, $v_2$ is measured up
to 10~GeV/$c$ and is still sizable above 8~GeV/$c$. The behavior of
$v_2$ at 62.4~GeV is similar to that observed at 200~GeV. It is argued
in~\cite{HighPt} that $v_2$ at large $p_t$ might
be related to the parton energy loss mechanism and may thus provide a constraint
on the initial gluon density.

The $p_t$ dependence of the charged particle $v_4$ is shown for both
energies in the right panel of Fig.~\ref{fig:v2_pt}. It is seen that
$v_4$ increases quadratically at low $p_t$ and has, like $v_2$, its
maximum around 3~GeV/$c$. At 200~GeV, $v_4$ is measured up to
7~GeV/$c$ and is still sizable above 6~GeV/$c$. Similar values are
obtained at 62.4~GeV.

\begin{figure}
\centering
\begin{minipage}[t]{77mm}
\includegraphics[width=1.0\textwidth]{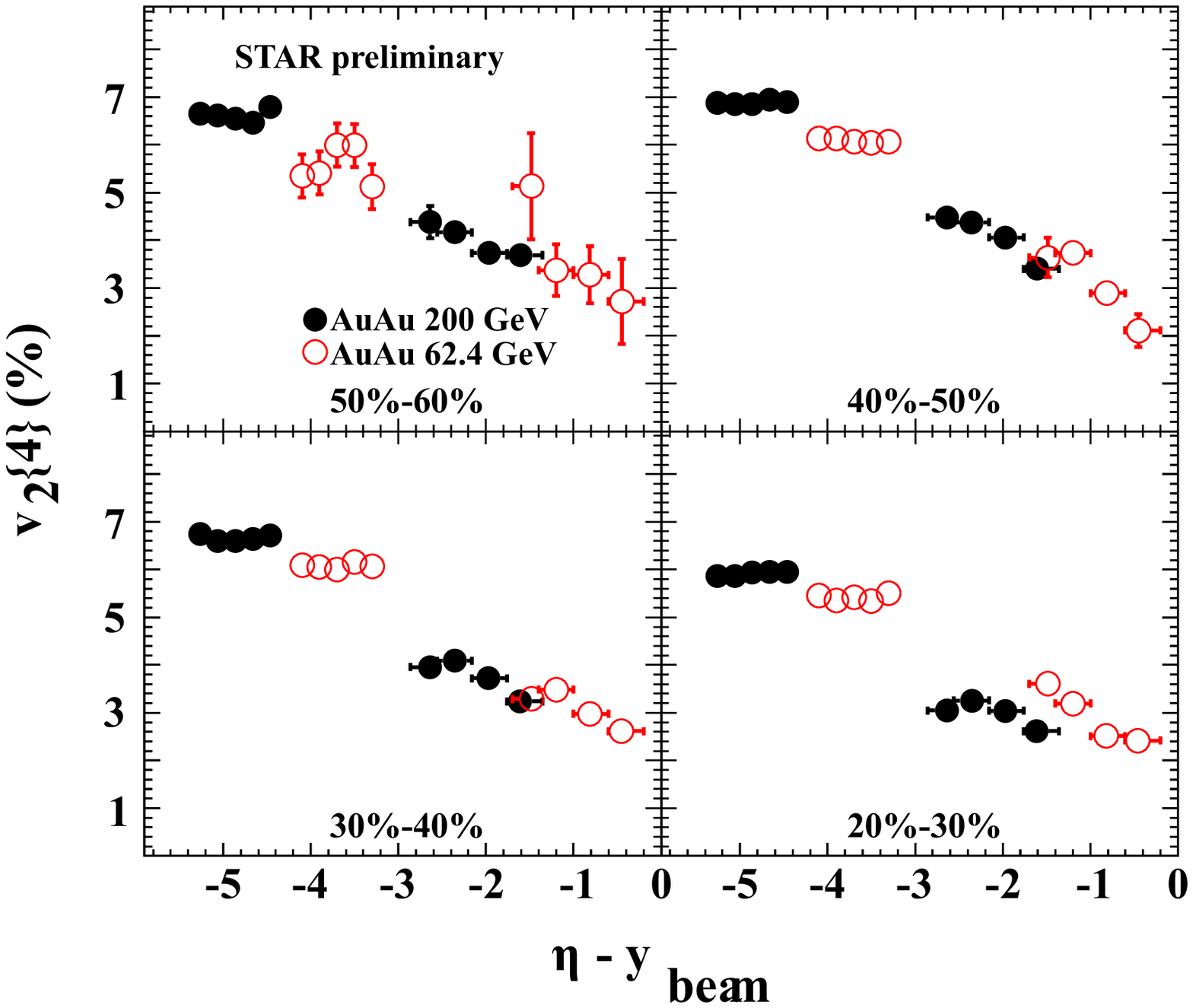}
\end{minipage}
\hfill
\begin{minipage}[t]{77mm}
\includegraphics[width=1.0\textwidth]{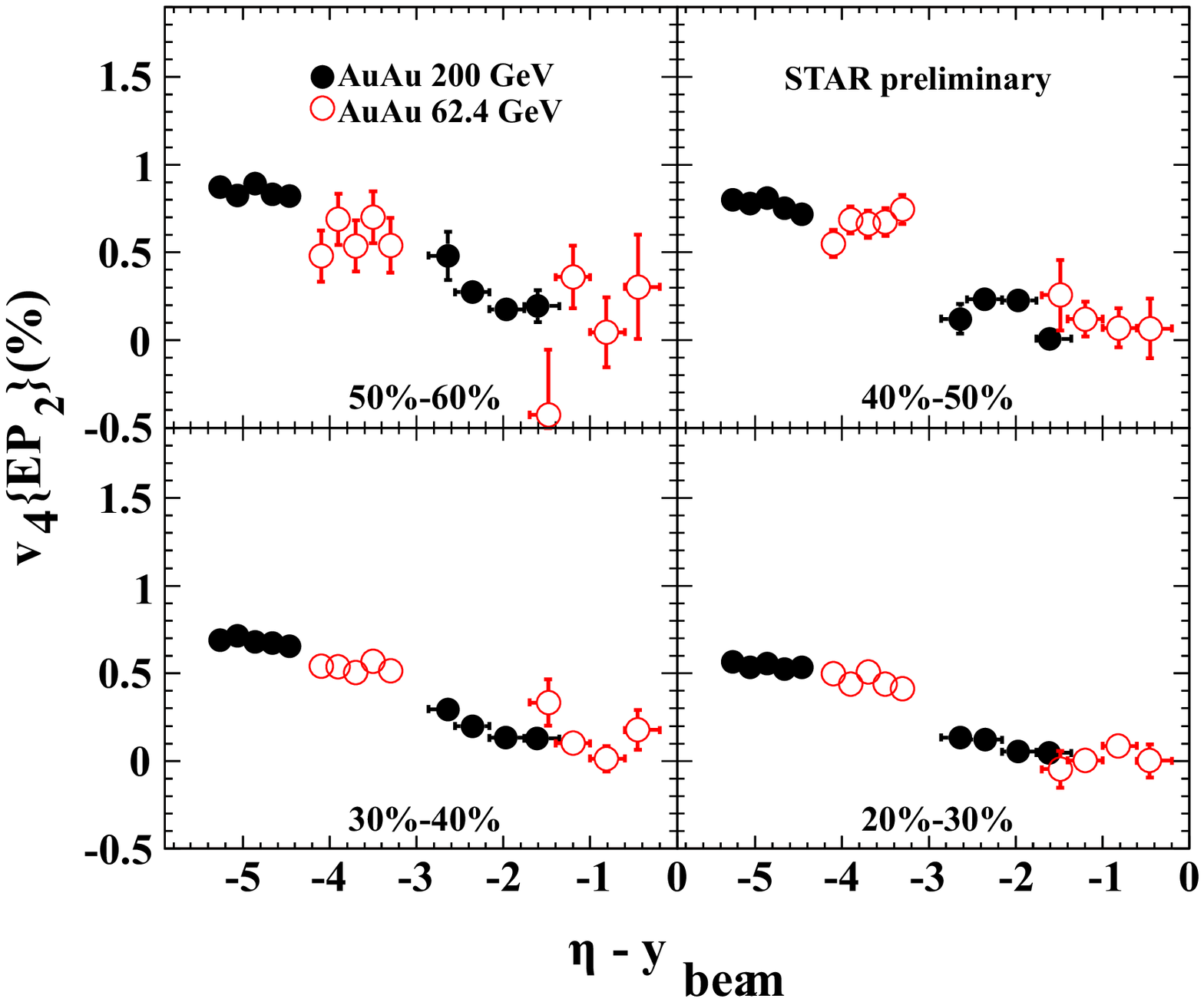}
\end{minipage}
\caption{Charged particle $v_2\{4\}$ and $v_4\{EP_2\}$ 
 versus $\eta - y_{beam}$ for
different centrality bins at 200 (full circles) and 62.4~GeV (open
circles). The flow coefficients are shown for particles in the forward hemisphere only.}
\label{fig:v2_eta}
\end{figure}

\paragraph{Rapidity dependence.} 
It has been shown that particle
production in the fragmentation region exhibits longitudinal scaling
when plotted as a function of $\eta - y_{beam}$~\cite{Phobos}. It is
also known that the integrated elliptic flow for fixed centrality at mid-rapidity 
is proportional to the particle yield $dN/dy$~\cite{Aihong}. 
If this scaling with $dN/dy$ holds at all
rapidities, then $v_2$ is also expected to show a longitudinal
scaling behavior. Figure~\ref{fig:v2_eta} (left) shows $v_2$ as a function of
$\eta - y_{beam}$ for different centralities at 200 and 62.4~GeV. The
$v_2$ values measured at both energies fall on a universal curve, indicating that 
the longitudinal scaling approximately holds. This scaling is also
observed for $v_4$ as can be seen in the right panel of
Fig.~\ref{fig:v2_eta}.

\begin{figure}
\centering
\begin{minipage}[t]{77mm}
\includegraphics[width=1.0\textwidth]{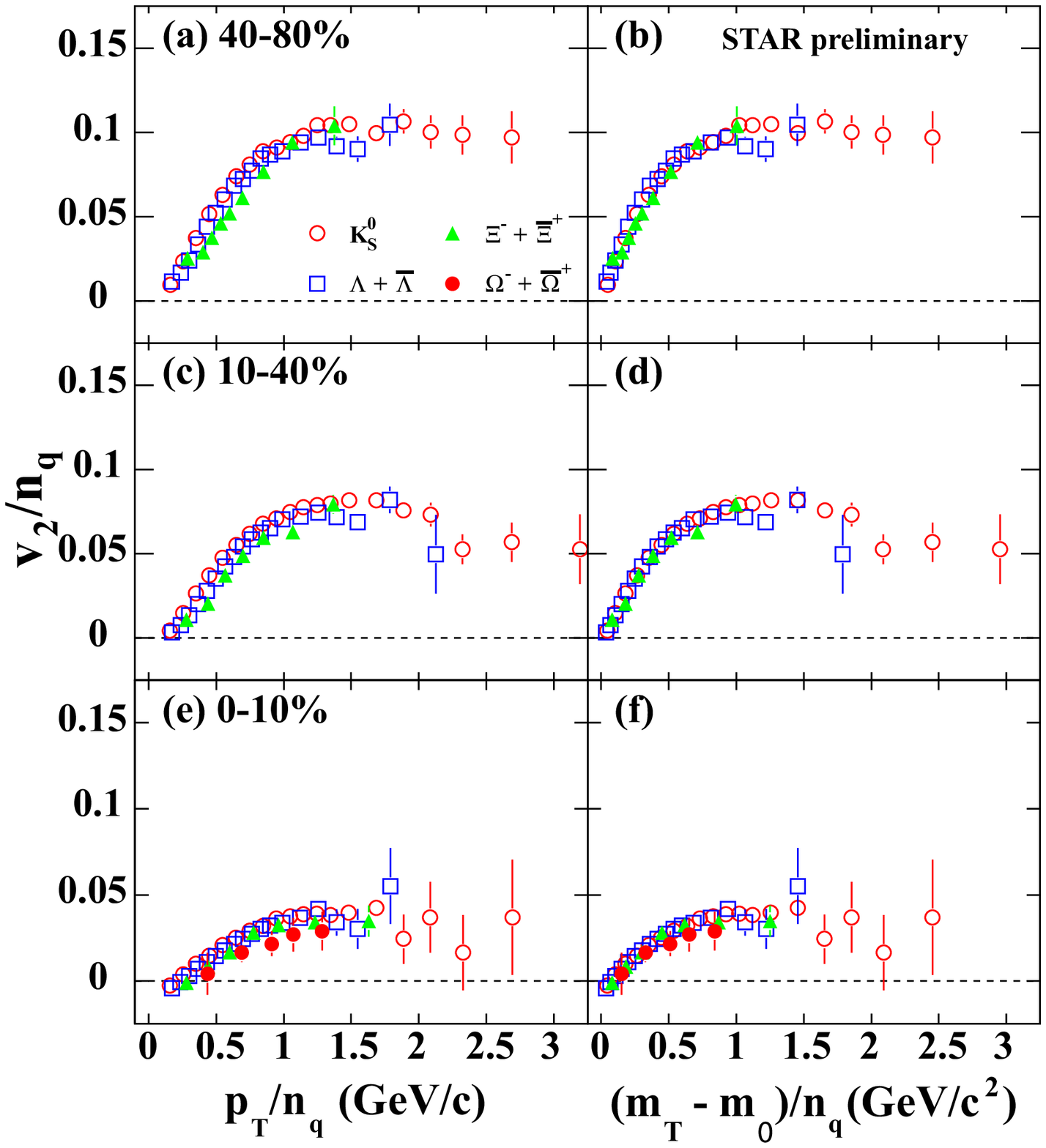}
\end{minipage}
\begin{minipage}[t]{77mm}
\includegraphics[width=1.0\textwidth]{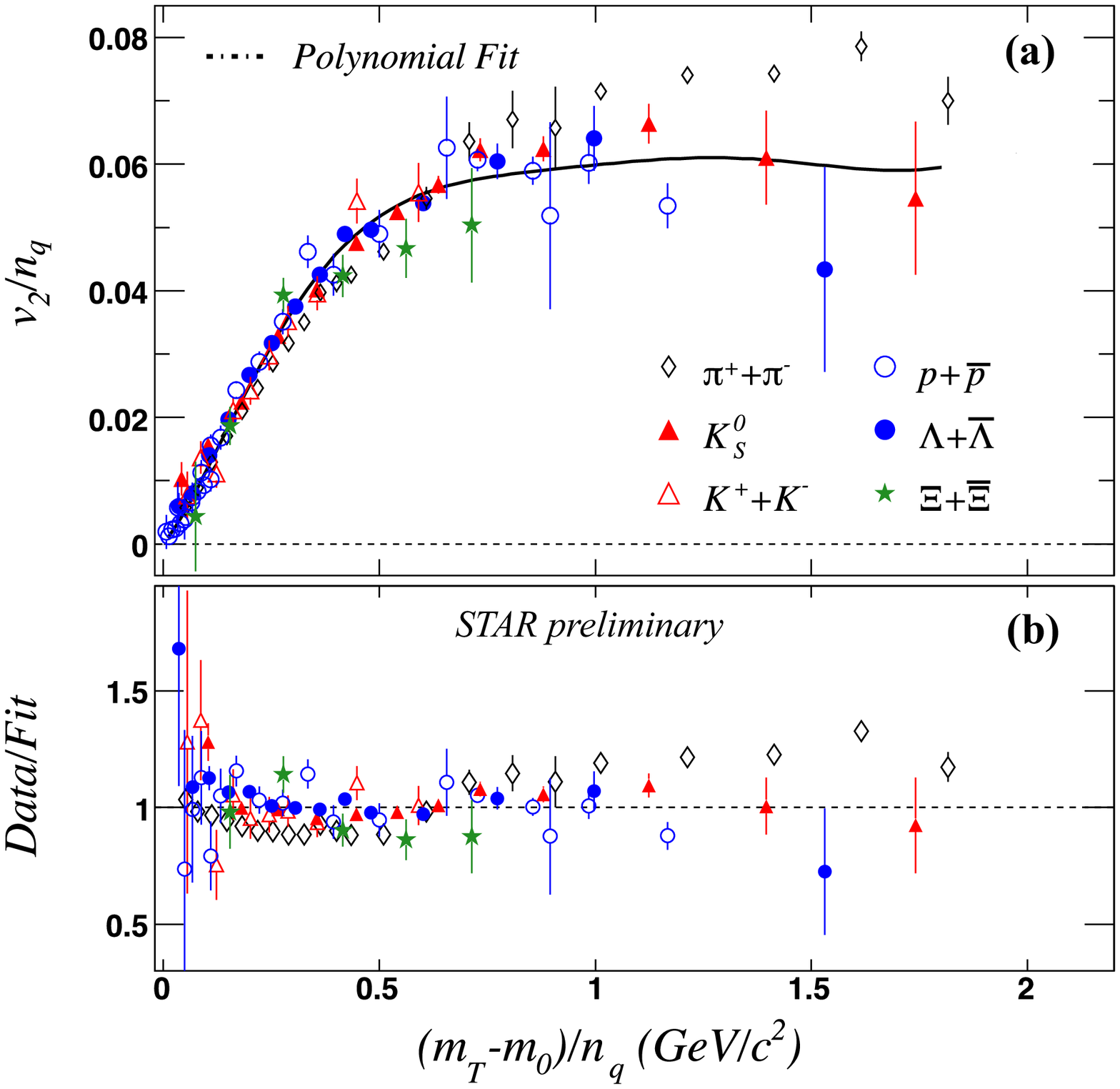}
\end{minipage}
\caption{The scaled flow coefficients $v_2/n_q$ versus $p_t/n_q$
 and $(m_T-m_0)/n_q$. Left: $v_2/n_q$ for 3 different centrality classes 
at $|\eta| < 1.0$ measured at 200~GeV. Right: $v_2/n_q$ versus $(m_T-m_0)/n_q$ 
for minbias data at 62.4~GeV for $|\eta| < 1.0$. The curve
shows a polynominal fit to the data.}
\label{fig:v2PID}
\end{figure}

\paragraph{Mass and particle type dependence.} 
It has been shown by STAR~\cite{StarPID} that $v_2$ for identified particles 
at low $p_t$ exhibits a mass ordering.
This ordering is well described by hydrodynamic
calculations which indicates that all particles flow with a common velocity. 
At intermediate $p_t$, $v_2$ scales with the number of constituent quarks
$n_q$~\cite{StarNQ}. This scaling can be explained in the coalescence picture
and is indicative of the partonic origin of flow~\cite{Molnar}. The ratio
$v_2/n_q$ is shown in Fig.~\ref{fig:v2PID} (left)
for three different centralities at 200~GeV as a function of $p_t/n_q$
and as a function of the scaled transverse mass $(m_T-m_0)/n_q$. The
scaled transverse mass, which takes into account relativistic effects,
is sometimes considered to be a better scaling
variable than $p_t/n_q$~\cite{Phenix}. 

An indication that $v_2$ follows constituent quark scaling independent of
centrality is given by the fact that the $v_2/n_q$ falls onto a
universal curve for each centrality bin. Figure~\ref{fig:v2PID} shows
that the constituent quark scaling holds at both 200~GeV and 62.4~GeV.

\paragraph{The ratio $v_4/v_2^2$} 
In recent work~\cite{OllitraultV4}, $v_4/v_2^2$ is proposed as a more
\begin{figure}[t]
\centering
\includegraphics[width=0.6\textwidth]{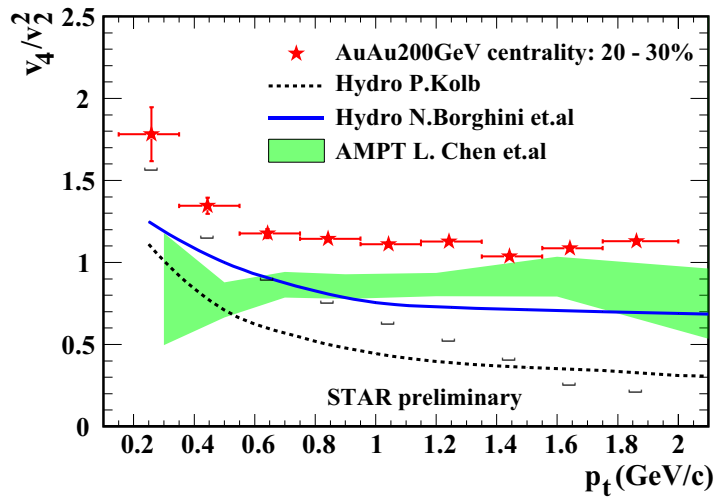}
\caption{The ratio of $v_4/v_2^2$ versus $p_t$ for charged particles 
at $|\eta| < 1.3$ in Au+ Au collisions at 200~GeV. The horizontal brackets indicate
the systematic uncertainty. The curves correspond to two hydrodynamic 
calculations~\cite{KolbV4,OllitraultV4}.
The filled area shows the AMPT model calculations~\cite{AMPT}.}
\label{fig:v4V2Sqr}
\end{figure}
sensitive probe of ideal hydrodynamic behavior. Furthermore, this
ratio is directly related to the degree of thermalization of the
medium, see also~\cite{Ko}. Under the assumption that $v_2\{4\}$ is a
genuine measure of elliptic flow, the systematic error in the ratio
$v_4/v_2^2$ is dominated by non-flow contributions to $v_4$. It can
be shown that the non-flow contribution to $v_4$ is proportional to
the difference $v_2\{2\}^2 - v_2\{4\}^2$. Obtaining this difference
from the data allows us to estimate the systematic error on
$v_4/v_2^2$. However, the difference can also originate from flow 
fluctuations~\cite{Fluctuation}. In that case, the systematic error will be reduced.
A detailed description of this systematic analysis is
beyond the scope of these proceedings and will be described in a
future publication.

Figure~\ref{fig:v4V2Sqr} shows the ratio $v_4/v_2^2$ measured at
200~GeV in the 20--30\% centrality interval. The horizontal brackets
in this figure show the lower limit of the systematic uncertainty as
presently estimated. In this figure, the data are compared to two
hydrodynamic model calculations~\cite{OllitraultV4,KolbV4} (curves)
and to model predictions based on a microscopic partonic and hadronic
description of the collision (AMPT model, filled area)~\cite{AMPT}.
It is seen that the data lie above the model predictions. However, the
present systematic uncertainties do not allow us to either validate or
exclude these two models.

\end{document}